# Formation of ferromagnetic bulk amorphous $Fe_{40}Ni_{40}P_{14}B_6$ alloys

Qiang Li

Department of Physics, The Chinese University of Hong Kong

Shatin, N.T., Hong Kong, P.R. China

**Abstract**

Ferromagnetic bulk amorphous $Fe_{40}Ni_{40}P_{14}B_6$ alloy rods with a diameter of ~ 1.2 mm can be prepared by means of a rapid quenching technique. If a fluxing technique is also used, amorphous rods with a diameter as large as ~2.5 mm can be synthesized. The critical cooling rate $R_c$ for the glass formation $Fe_{40}Ni_{40}P_{14}B_6$ is estimated to be on the order of $10^2$ $K\ s^{-1}$.

**Introduction**

Conventionally, a metallic glass is prepared by rapidly quenching its melt from above its liquidus, $T_l$, to below its glass transition temperature $T_g$ [1]. Typical quenching rates are on the order of $\sim 10^6$ K s$^{-1}$. As a result, one dimension of the as-prepared specimen is very thin, about 30 – 50 μm, to facilitate heat removal [2]. The small physical size has so far limited industrial/commercial applications of this class of materials.

Driven by the potential applications of amorphous alloys, the search for bulk amorphous metals has persisted for the past two decades. By 'bulk', it means that an as-synthesized specimen has a dimension ≥ 1 mm measured in all directions. By a fluxing technique, Kui et al.[3] were able to prepare bulk amorphous $Pd_{40}Ni_{40}P_{20}$ ingots with a diameter of ~1 cm. Since then extensive and important progresses were made in Inoue's group. Mg- [4], Ln- [5], Zr- [6,7], Fe- [8,9], Pd-Cu- [10], Ti- [11,12], and Ni- [13] based bulk glasses were successively manufactured by a casting method. Built on Inoue's work, Peker and Johnson [14] discovered the formation of bulk metallic glass $(Zr_3Ti)_{0.55}(Cu_5Ni_4)_{0.225}Be_{0.225}$ which has a trade name of Vitreloy 1. There are recent reviews [15,16] on the formation of bulk amorphous metallic alloys.

Perhaps the most attractive aspect of amorphous alloys is their soft magnetic behavior. They can be used in electrical transformers, motors, and high frequency switching supplies [17]. Before the emergence of Fe-based ferromagnetic bulk glasses [8,9,18], quite a number of ferromagnetic alloys that form glass less readily were already found. $Fe_{40}Ni_{40}P_{14}B_6$ is one of them and is the first commercial product of Allied Chemical

Corporation under the trade name METGLAS 2826. It has excellent soft magnetic properties, but small physical sizes, typically as 50 μm thick ribbons. In this Rapid Communication, we report the bulk glass formation of $Fe_{40}Ni_{40}P_{14}B_6$ with the critical cooling rate of ~$10^2$ K s$^{-1}$, by means of a water-quenching method.

## I. Experimental Procedures

$Fe_{40}Ni_{40}P_{14}B_6$ ingots were prepared from Fe chips (99.98% pure), Ni spheres (99.95% pure), B pieces (99% pure), and $Ni_2P$ ingots (The $Ni_2P$ ingots used were themselves prepared from powders (98% pure)). After the right proportion was weighed, they were put in a clean fused silica tube and alloying was brought about in a rf induction furnace under Ar atmosphere. All the as-prepared specimens had a mass of ~3 g.

Bulk ferromagnetic Fe-Ni-P-B glasses were prepared via two different routes. The first one involves direct quenching of an untreated $Fe_{40}Ni_{40}P_{14}B_6$ melt. In the second route, a $Fe_{40}Ni_{40}P_{14}B_6$ melt that had been fluxed [3] was subjected to water-quenching. In the fluxing technique, a molten alloy was immersed in a molten oxide at an elevated temperature for a prolonged period. The oxide or fluxing agent served to remove impurities from the molten specimen.

Water-quenching of a molten alloy was conducted in a drawn fused silica tube as shown in Fig. 1. It consists of two fused silica tubes of different diameters connected to each other. Such a system is called J in the followings. Typical dimensions of larger tubes in J are: length, 5-10 cm; and inner/outer diameter, 11/13 mm. Those for smaller tubes are: length, 5-10 cm; wall thickness, 0.1~0.2 mm; and inner diameter, 1-3 mm.

In the experiment, a $Fe_{40}Ni_{40}P_{14}B_6$ ingot, fluxed or without fluxed, was put in J. It would sit on and block the open end of the smaller tube. J was then connected to a mechanical pump that evacuated it to $\sim 5\times10^{-3}$ torr. Next it was filled out high purity Ar gas of just below 1 atm in the whole tube. At this point, a torch was used to melt the ingot. To achieve rapid quenching, the melt was pushed into the smaller fused tube, where the wall thickness was thin, by introducing Ar gas into the system through the larger tube. After this, the whole system was immediately transferred to a furnace that heated the melt to ~1450 K, well above the liquidus of the $Fe_{40}Ni_{40}P_{14}B_6$ alloy, ~1200 K. The system sat in the furnace for about 5 minutes before removed from the furnace and quenched in salted ice-water. An as-prepared specimen was in rod form with a diameter of d, which is also the inner diameter of the smaller tube in J.

The amorphization of the as-formed rod specimens was checked by X-ray diffractometer with Cu K$\alpha$ radiation and transmission electron microscopy (TEM). The thermal behavior of as-prepared specimens was examined at a heating rate of 0.33 K s$^{-1}$ by differential scanning calorimetry (DSC) and differential thermal analysis (DTA). Thermomagnetic measurement was performed at a heating rate of 0.33 K s$^{-1}$ by thermal gravitation analysis (TGA) combined with a magnet that produced a field of about 50 Gauss at the position of specimen. Magnetic properties of specimens at 300 K were measured in a maximum of applied field of 1.0 T at a scan rate of 0.1 T/min by means of an Oxford Maglab VSM with the superconducting magnet and the mass of specimens is about 30 ~ 50 mg.

## II.     Results

Two kinds of ingots were used to prepare the rod specimens, one fluxed and the other not. For clarity, we first focus on rod specimens that were prepared from fluxed ingots. All as-prepared rods had shiny surface. In order to determine the uniformity of their microstructure further, these as-prepared rods were etched in a solution of $HNO_3:HCl:H_2O$ = 1:1:3 in a supersonic cleaner for about ten minutes. It turned out that for rod specimens with d ≤ 1.5 mm, a rod with the length exceeding 10 cm could be prepared and the whole rod surface can remain shiny during etching. On the other hand, for d ≥ 1.5 mm, frequently both ends of the etched rods became gray and only the middle part with the length varying from 3 to 8 cm still remained shiny surface. For d > 2.5 mm, no such rods that could remained shiny in every part of their surface after being etched, could be prepared. Some of the part, which surface can remain shiny in the period of etching, of the as-prepared rods are shown in Fig. 2. By means of examination of DSC, it was found that the thermal scan of those shiny parts of the as-prepared rods after being etched exhibited a glass transition but that of the gray parts don't. Only the part, which surface remained shiny after being etched, of the as-prepared rods are studied in this paper and the "specimen" in the following section is just pointed to these parts.

X-ray analysis was used to characterize the specimens and an as-obtained diffraction pattern is shown in Fig. 3 that depicts a broad peak, illustrating the amorphous nature of the specimen. The raised baseline is due to fluorescent scattering of iron and nickel atoms for Cu Kα radiation. A specimen was then subjected to TEM studied. A bright-field image together with its diffraction pattern and corresponding dark-field image are shown in Fig. 4 (a) and (b) respectively. Both bright-field and dark-field images show a homogeneous background while the diffraction pattern consists of halos. These confirm that the specimen is a homogeneous glass. Finally, a typical DSC thermal scan of the as-prepared rods at a heating rate of 0.33 K s$^{-1}$ is shown in Fig. 5. The glass transition $T_g$ and the kinetic crystallization temperature $T_x$ can be determined to be 659.0 K and 689.8 K respectively from this thermal scan.

The magnetic properties of these ferromagnetic bulk glasses were studied by means of VSM. The measured saturation magnetization $B_s$ is 92.3 emu/g. The density of bulk $Fe_{40}Ni_{40}P_{14}B_6$ amorphous rods can be determined to be 7.41 kg/m$^3$ by means of immersed water method and thus the saturation magnetization $B_s$ of specimens can be deduced to be 0.859 T. Unfortunately, coercivity $H_c$ of the specimen could not be resolved by the instrument. The Curie temperature $T_c$ of the amorphous alloy was determined to be 532.7 K by a thermomagnetic technique with a heating rate of 0.33 K s$^{-1}$.

In the second case, ingots that were not fluxed in advance were subjected to rapid quenching in J. For $1 \leq d \leq 1.2$ mm, only the middle part surface of as-prepared rod could remain shiny after being etched. The lengths of this central portion can stretch from 3 cm to 6 cm for our experimental arrangement described earlier. For $d > 1.2$ mm, there is no any parts in the whole rod that can remain its shiny surface. X-ray analysis, TEM and DSC confirm that the shiny part of a rod is amorphous, similar to that in the first case. Moreover, they exhibit similar magnetic properties.

IV. Discussion

In the previous section, it is pointed out that crystallization frequently occurred in both the ends of rod in both the routines. This can be understood as the following reasons. After molten specimen was injected into J tube, there are free surfaces existed in the both ends of sample. The oxides will be formed on these surfaces during high temperature isothermal proceeding. These oxides with high melting point can become the efficient heterogeneous nucleation sites and have a chance to induce the crystallization of sample at lower cooling rate. Thus it is required larger critical cooling rate in the both ends of rod than that in the middle of rod.

The reduced glass transition temperature of $Fe_{40}Ni_{40}P_{14}B_6$ is 0.57 and it implies moderate glass formability. In general it is considered that the critical cooling rate $R_c$ for the glass formation of $Fe_{40}Ni_{40}P_{14}B_6$ is on the order of $\sim 10^5$ K s$^{-1}$ [15]. In 1993 Diefenbach et al. [20] reported to prepare the amorphous $Fe_{40}Ni_{40}P_{14}B_6$ alloy sphere in diameter up to 300μm by means of the containerless solidification technique and corresponding critical cooling rate is estimated to be 8000 K s$^{-1}$. Most recently, Shen et al. [21] employed the flux-melting and water-quenching technique to prepare bulk glassy $Fe_{40}Ni_{40}P_{14}B_6$ alloys in the form of 2-mm diameter spheres and 1-mm diameter rods. It is indicated that the heterophase impurities has been removed and restricted more properly in our technique. It results in the maximum diameters of bulk amorphous $Fe_{40}Ni_{40}P_{14}B_6$ alloy rods prepared from the fluxed specimen is ~2.5 mm and the corresponding critical cooling rate is estimated to be on the order of $\sim 10^2$ K s$^{-1}$. And for un-fluxed specimen, the maximum diameter of bulk amorphous alloy rods is ~1.2 mm. It is pointed out again that fluxing is an effective technique to remove the impurities in the alloy system. Meantime it should also be noted that the bulk amorphous $Fe_{40}Ni_{40}P_{14}B_6$ alloy rod with a diameter as large as ~1.2 mm can be prepared from the as-prepared $Fe_{40}Ni_{40}P_{14}B_6$ ingot without any further treatments in our experiment. Compare with the other researcher's work, it seems that the inert quartz tube employed in our technique has played an important role. As mentioned above, crystallization always starts at the two ends of a rod specimen in our experiment. This is different from that in the copper mold casting technique performed by A. Inoue's group in which the crystallization usually starts at the interface between the specimen and the copper mold wall [22]. It means that the impurities in the free surfaces at the two ends of the molten rod are more potential nuclei than the interface between the molten specimen and the quartz tube wall in our experiment. It is pointed out that the inert quartz tube is important for preparing the bulk amorphous alloys in our experiment although its poor thermal conductivity limits the cooling rate of the specimen. Thus it is indicated that the inert container is another important factor in addition to the large cooling rate and the proper elimination of impurities in preparation of bulk amorphous alloys. Therefore the critical cooling rate of $Fe_{40}Ni_{40}P_{14}B_6$ was improved further in this study by the flexible experiment setup. This result implies that if heterogeneous impurities are properly removed and controlled, the Fe-based systems discovered by Inoue at al. [8,9] may probably be melt-quenched to very large rods, with a diameter > 1 cm, a size that renders them to direct industrial applications.

Thermal and magnetic properties of as-prepared bulk amorphous $Fe_{40}Ni_{40}P_{14}B_6$ rods determined in this work are summarized in Table 1 and corresponding data obtained from the thin $Fe_{40}Ni_{40}P_{14}B_6$ glassy ribbons reported by other authors are also included in Table 1 for comparison purposes. From Table 1 it can be indicated that $T_g$ and $T_x$ of our specimen have no significantly differences from that of thin $Fe_{40}Ni_{40}P_{14}B_6$ glassy ribbons prepared with a cooling rate of $1\times10^5$ K s$^{-1}$. It seems that the thermal characteristics of amorphous alloy are not influenced significantly by its preparation history. In fact such results had also been found in Pd-Ni-P amorphous alloy [23]. As we know, the glass-transition temperature of a particular material is not an intrinsic property and depends on its thermal history and it is found that the slower is the cooling rate, the lower is the glass-transition temperature [24]. It seems that our experimental results do not lead to this conclusion. However we should not make such a conclusion because, if the $T_g$ of a glass is determined by heating the material, the previous result may be expected unless the heating rate is equal to the original cooling rate [24]. For metallic glass, such a measurement is difficult to be performed due to its high critical cooling rate. Thus the glass transition temperature $T_g$ obtained at a very low heating rate relative to its original cooling rate is determined only by the thermodynamic properties of amorphous alloy, i.e. the active energy associated with atom diffusion, which may be little influenced by the kinetic history for metallic glass. However it should be noted that the enthalpy of crystallization of our specimen is reasonably smaller than that obtained from the thin glassy ribbons and this parameter indicates the residual entropy of the specimen relative to its ideal glass state, which entropy is considered to be the same as that of the corresponding crystalline phase. Therefore it can be concluded that the arrangement of atoms in our specimen is closer to the ideal glass state than that of the thin glassy ribbons prepared at a cooling rate of ~$10^5$ K s$^{-1}$.

The values of $B_s$ and $T_c$ of our specimens are comparable, but reasonably larger to those obtained from thin $Fe_{40}Ni_{40}P_{14}B_6$ glassy ribbons prepared at a cooling rate of $1\times10^5$ K $s^{-1}$. Libermann, Graham and Flanders [25] have shown that $T_c$ is sensitive to the degree of relaxation, rising when the as-prepared glass is annealed. Thus the improvement of $T_c$ may imply that our specimen prepared at a low cooling rate result in more relaxed structure than that of specimen prepared by means of melt-spinning technique. This conclusion is in agreement with the above section. The increase in $B_s$ is probably related to the difference of the short-range ordering of atom in the specimens prepared by different methods.

### V. Conclusion

(1) Ferromagnetic bulk amorphous $Fe_{40}Ni_{40}P_{14}B_6$ alloys rods with a diameter of ~1.2 mm can be prepared by means of a rapid quenching technique. If a fluxing technique is also used, amorphous rods with a diameter as large as ~2.5 mm can be synthesized and the corresponding critical cooling rate $R_c$ for the glass formation of $Fe_{40}Ni_{40}P_{14}B_6$ is estimated to be on the order of $10^2$ K $s^{-1}$. The improvement of the critical cooling rate should be owed to efficient elimination of heterogeneous impurities in the specimens and the usage of the inert quartz tube in our experiment.

(2) The thermal stability and magnetic properties of bulk $Fe_{40}Ni_{40}P_{14}B_6$ amorphous rods are comparable to those obtained from thin $Fe_{40}Ni_{40}P_{14}B_6$ glassy ribbons prepared at a cooling rate of $10^5$ K $s^{-1}$. Small differences between them can be understood by the reason that the atom arrangement of bulk $Fe_{40}Ni_{40}P_{14}B_6$ amorphous specimens prepared at a cooling rate of ~$10^2$ K $s^{-1}$ in our work is closer to the ideal glass state than that of the specimens prepared by means of melt-spinning technique.

**Acknowledgement**

We thank Hong Kong Research Grants Council for financial support.

**Table 1**     Summary of thermal and soft magnetic properties of bulk rods and glassy ribbons reported by other authors

| Reference | Thermal properties | | | | | Magnetic properties | | |
|---|---|---|---|---|---|---|---|---|
| | $T_g$ | $T_x$ | $\Delta H_x$ | $\Delta T_x$ | $T_g/T_m$ | $T_c$ | $H_c$ | $M_s$ |
| | K | K | kJ/mol | K | | K | A/m | T |
| **This work** | 662.1[a] | 694.9[a] | 4.98[a] | 32.8[a] | 0.56[b] | 532.7[c] | - | 0.859 |
| [26]* | 658[a] | 694[a] | 5.51[a] | 36[a] | - | 496[a] | - | - |
| [27]-C* | - | - | - | - | - | 520[d] | | 0.82 |
| [27]-K* | - | - | - | - | - | 523[d] | | 0.83 |

\* The specimens were prepared by means of melt-spinning technique.

(a) Determined by means of DSC at a heating rate of 0.5 K s$^{-1}$;

(b) $T_m$ was determined to be 1184.2 K by means of DTA at a heating rate of 0.33 K s$^{-1}$;

(c) Determined by means of thermomagnetic method with a heating rate of 0.33 K s$^{-1}$;

(d) The heating rate is unknown.

**Figure Captions**

Fig. 1  Schematic diagram of the experimental setup, J, made up of two fused silica tube of different diameters, sits in the middle of the furnace.

Fig. 2  Some as-prepared ferromagnetic bulk $Fe_{40}Ni_{40}P_{14}B_6$ alloy rods are displayed.

Fig. 3  X-ray diffraction pattern of an amorphous $Fe_{40}Ni_{40}P_{14}B_6$ alloy. It is displays a broad peak, characteristic of amorphous materials.

Fig. 4  TEM studies of a bulk $Fe_{40}Ni_{40}P_{14}B_6$ amorphous alloy. (a) A TEM bright-field image displays a homogeneous background; (b) The corresponding electron diffraction pattern of the specimen shown in (a). There are halos only.

Fig. 5  A DSC scan of a bulk $Fe_{40}Ni_{40}P_{14}B_6$ amorphous alloy. The heating rate used is 0.33 $K\ s^{-1}$.

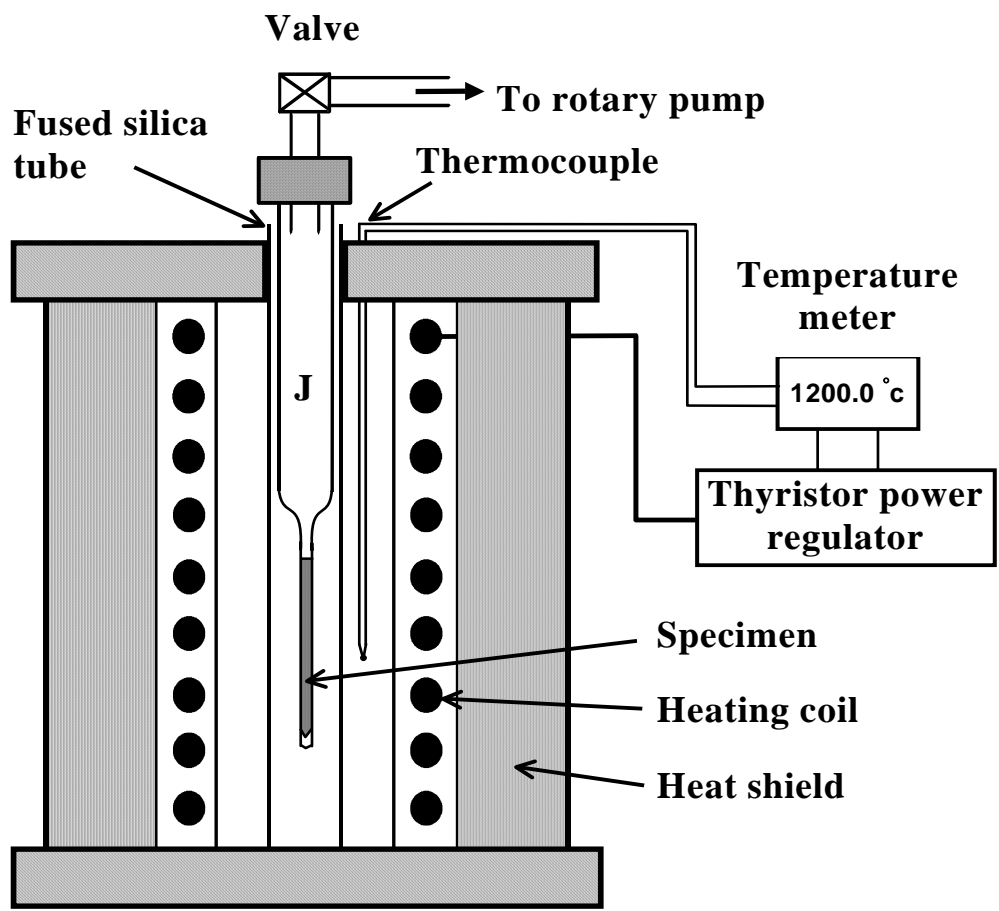

Fig.1 Schematic diagram of the experimental setup, J, made up of two fused silica tube of different diameters, sits in the middle of the furnace.

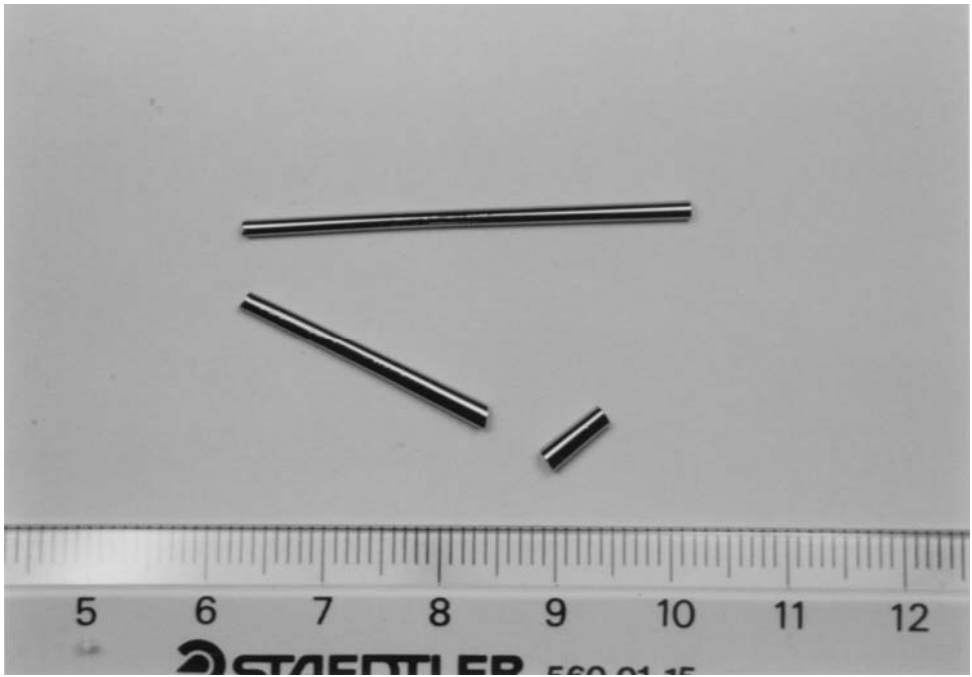

Fig.2 Some as-prepared ferromagnetic bulk $Fe_{40}Ni_{40}P_{14}B_6$ alloy rods are displayed.

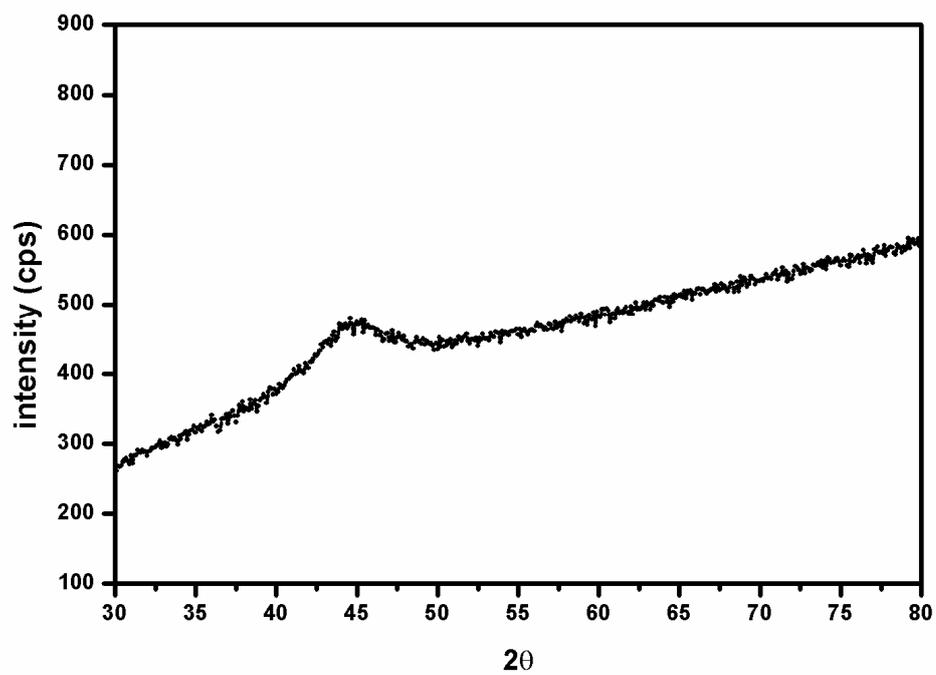

Fig.3 X-ray diffraction pattern of an amorphous $Fe_{40}Ni_{40}P_{14}B_6$ alloy. It is displays a broad peak, characteristic of amorphous materials.

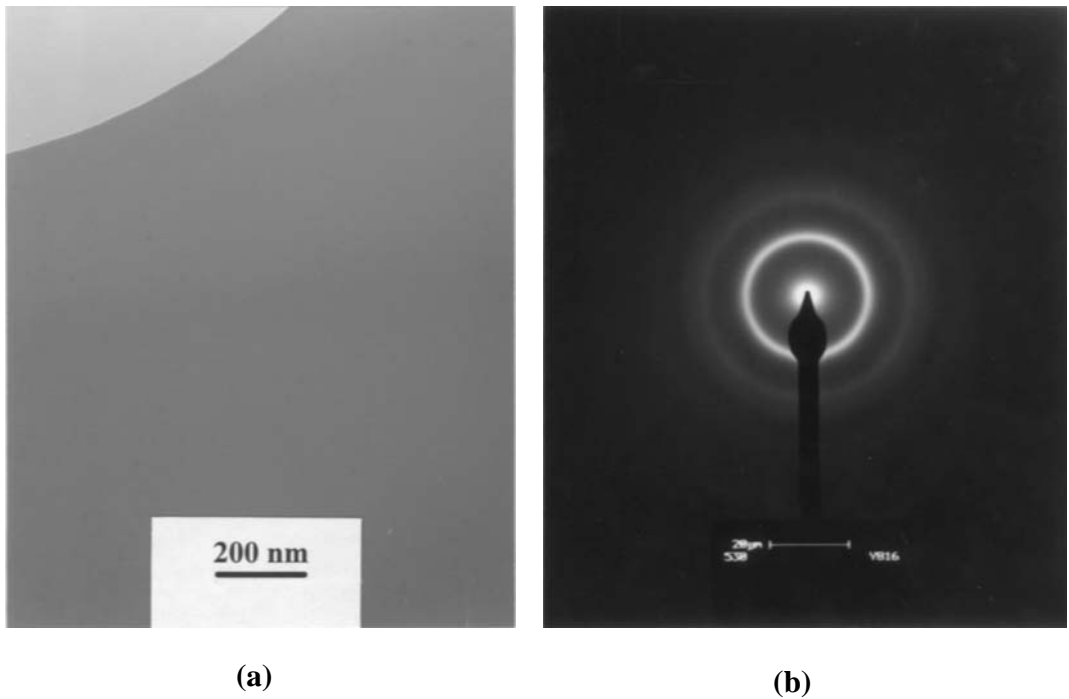

Fig.4 TEM studies of a bulk $Fe_{40}Ni_{40}P_{14}B_6$ amorphous alloy. (a) A TEM bright-field image displays a homogeneous background; (b) The corresponding electron diffraction pattern of the specimen shown in (a). There are halos only.

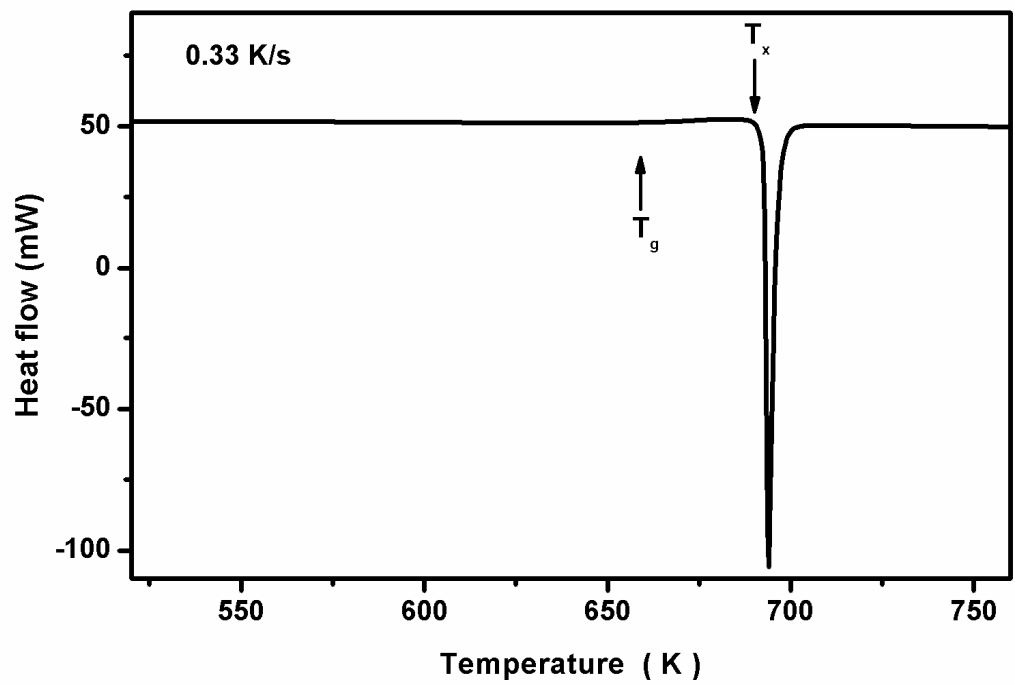

Fig.5  A DSC scan of a bulk $Fe_{40}Ni_{40}P_{14}B_6$ amorphous alloy. The heating rate used is 0.33 K s$^{-1}$.